# Exploring the Role of Gamification in Enhancing Academic Library Services: A Survey of Library Leaders in India

**Mr. Subaveerapandiyan A (Corresponding Author)**
Manager (Planning & Coordination)—Library
Vinayaka Mission's Research Foundation[a]
(Deemed to be University), Salem - 636 308, Tamil Nadu, India

Research Fellow
Department of Library
Shinawatra University[b]
Pathum Thani 12160, Thailand
Email: subaveerapandiyan@gmail.com
ORCID: https://orcid.org/0000-0002-2149-9897

**Ms. Pragya Lohia**
PhD Research Scholar
Department of Library and Information Science
Central University of Punjab, Bathinda, India
Email: prgylh99@gmail.com
ORCiD: https://orcid.org/0000-0003-4519-2993

**Dr. Dattatraya Kalbande**
Librarian
J.Watumull Sadhubella Girls College, Ulhasnagar, Maharashtra, India
Email: kalbanded@gmail.com
ORCID: https://orcid.org/0000-0002-5276-0453

**Dr Naved Ahmad (Corresponding Author)**
Assistant Professor
Department of Computer Science and Information Systems
College of Applied Sciences, AlMaarefa University
P.O.Box 71666, Riyadh, 13713 Saudi Arabia
Email: nahmad@um.edu.sa
ORCiD: https://orcid.org/0000-0002-4932-401X

**Mr. Kailash Chand Sharma**
PhD Research Scholar

Department of Library and Information Science
Central University of Punjab, Bathinda, India
Email: kk9875266830@gmail.com
ORCiD: https://orcid.org/0009-0003-2489-6112

*Abstract*
*This study explores the role of gamification in enhancing academic library services in India by surveying library leaders across various institutions. Using game-like elements in non-game contexts, gamification can boost user engagement and improve services such as information literacy and research consultations. Findings reveal moderate awareness and generally positive perceptions of gamification's effectiveness. However, challenges like insufficient staff expertise, infrastructure, and limited funding hinder implementation. The study emphasises the need for additional resources, including staff training and technological upgrades, to unlock the full potential of gamification in academic libraries.*

*Introduction*
Gamification, integrating game-design elements into non-game contexts, has emerged as a powerful tool to enhance engagement, motivation, and learning outcomes across various fields. Originally popularised in business and education, gamification involves applying mechanisms such as points, badges, leaderboards, and challenges to drive user participation and enjoyment. In educational settings, gamification has increased student motivation, improved learning experiences, and fostered active participation.

In academic libraries, traditionally seen as static and utilitarian environments, gamification presents an opportunity to revitalise the user experience. By transforming routine library activities into engaging and interactive experiences, libraries can enhance user engagement, promote resource utilisation, and support information literacy. Gamification can make library services more appealing and relevant, mainly when digital and interactive experiences are increasingly prevalent among students and faculty.

*Problem Statement*
Despite the potential benefits of gaming, many academic libraries face challenges in fully implementing and leveraging these strategies. The traditional approach to library services often results in limited user engagement and underutilisation of available resources. Libraries struggle to attract frequent interactions and effectively communicate their services' value to users. Moreover, there is often resistance to adopting new methods and technologies, which can hinder the integration of gamification into existing systems.

Current literature suggests that while gamification has been applied successfully in some library settings, comprehensive research on its effectiveness is lacking across diverse academic institutions. Issues such as varying levels of awareness, differing perceptions of effectiveness, and the challenges of integrating gamification into traditional library practices remain underexplored. Furthermore, the specific resources and support required for successful implementation have not been well-documented.

*Research Objectives*

This study aims to address these gaps by exploring the role of gamification in enhancing academic library services through the following objectives:
- To evaluate the level of awareness among library leaders about gamification and its application in academic libraries.
- To understand library leaders' perceptions regarding gamification's effectiveness in enhancing library services and user engagement.
- To determine which library services could benefit most from gamification.
- To identify the main challenges and barriers libraries face in implementing gamification strategies.
- To explore the resources and support needed for gamification's successful implementation and sustainability in academic libraries.

*Literature Review*

*Gamification in Library Services*

Gamification has been widely recognized as an effective strategy for enhancing user engagement in library settings (Brigham, 2015). Research suggests that incorporating game mechanics, such as rewards, challenges, and progress tracking, significantly increases user interaction with library services (Alaban & Singh, 2024; Brown & Kasper, 2013; Folmar, 2015).

Several gamification tools have been successfully implemented in academic libraries. Mozilla Open Badges has been used in reading programs to reward user participation, allowing learners to earn digital badges for completing specific learning tasks (Jug, 2023). The Lemontree system, implemented at Huddersfield University Library, encourages students to earn points and badges through regular library activities such as borrowing materials, renewing books, and engaging with digital resources (Walsh, 2015).

Another widely used tool is Kahoot!, an interactive quiz platform that enables students to engage in gamified learning activities. Aibar-Almazán et al. (2024) found that Kahoot! improved essential academic skills such as attention, creativity, and critical thinking when integrated into educational environments. Similarly, ClassDojo, a classroom behaviour management app, has been incorporated into university learning environments to enhance student motivation and collaboration through gamified participation and feedback mechanisms (Latorre-Cosculluela et al., 2025).

Academic libraries have also integrated gamification into their orientation programs. The Mobile Scavenger Hunt, developed by North Carolina State University Library, helps new students navigate library resources through an engaging, game-based experience (Adeyemi et al., 2021). Similarly, digital escape rooms and interactive quests have introduced first-year students to research databases and citation management tools (Saikia et al., 2023).

Another noteworthy example is FIRE (First-year Information Research Experience), an online gamification initiative introduced by the University of New Hampshire Library to enhance students' research skills and familiarize them with information literacy practices (Khademi Zare et al., 2024). Additionally, the Murder at the Met interactive game, designed by the Metropolitan

Museum of Art, has been adapted by academic libraries to engage users in research-based problem-solving tasks (Garaccione et al., 2024).

Furthermore, gamification can create a more immersive and enjoyable experience for users, leading to an increased frequency of library visits and utilisation of services (Kitamura & Sumi, 2023; Reed & Miller, 2020). A survey revealed that libraries implementing gamified programs experienced a notable increase in user participation and interaction with library resources. The study highlighted that gamification strategies, such as scavenger hunts and interactive quizzes, effectively captured users' interest and encouraged them to engage with library materials more meaningfully (Kitamura & Sumi, 2023; Zhou & Yang, 2023).

Gamification has also been shown to enhance library users' information literacy and research skills (Jansukpum et al., 2024). By incorporating game elements into information literacy programs, libraries can create engaging and interactive learning experiences that help users develop critical research skills (Young, 2016). A study by Encheva et al. (2020) demonstrated that gamified information literacy tutorials improved students' understanding of research concepts and ability to locate and evaluate sources. Game-based scenarios and interactive exercises facilitated active learning and reinforced key information literacy competencies (Pinto et al., 2024).

Gamification can help users retain information more effectively (Cai et al., 2024). Research by Riedmann et al. (2024) suggests that the motivational aspects of gamification, such as immediate feedback and rewards, can enhance cognitive engagement and memory retention. Libraries integrating these elements into their information literacy programs can improve users' research skills and overall information literacy (Withorn et al., 2020, 2021).

By integrating game elements into library services, institutions can create a more dynamic learning environment, ultimately enhancing user motivation and participation in academic research and literacy programs.

*Challenges and Limitations*
Despite its benefits, the implementation of gamification in libraries is not without challenges. One significant challenge is the technical complexity of designing and integrating gamified elements into library systems. Libraries may face difficulties developing and maintaining digital platforms supporting game mechanics, such as leaderboards and progress tracking (Alaban & Singh, 2024; Brigham, 2015; Zeybek & Saygı, 2024). According to studies by Sánchez-Mena and Martí-Parreño (2017) and van Roy and Zaman (2017), technical issues related to software compatibility and user interface design can hinder the successful implementation of gamification strategies.

Resistance to change is another common challenge. Some staff and users may be sceptical about the effectiveness of gamification and resist adopting new approaches (Sanchez et al., 2020). Research by Alaban and Singh (2024) and Dalili Saleh et al. (2022) highlights that addressing concerns and providing adequate training and support can help mitigate resistance and facilitate the successful integration of gamification in library services. Gamification platforms are crucial for incorporating game mechanics like leaderboards, point systems, and badges, which are essential to a successful gamification strategy (Eden, 2015; Folmar, 2015; Lim, 2023).

Some critics argue that gamification, mainly when focused on extrinsic rewards like points and badges, may undermine intrinsic motivation for learning and exploration (Alsawaier, 2018; Richter et al., 2015). This concern is especially relevant in academic libraries, which aim to foster deep learning and critical thinking rather than just task completion. Additionally, gamified systems that rely on social comparisons, such as leaderboards, may discourage lower-performing users or create unhealthy competition among patrons (Appleton, 2020; Kabilan et al., 2023).

### *User Perceptions and Preferences*

Gamification has gained popularity as a method to increase user engagement in libraries by incorporating elements of fun, competition, and interaction. However, user perceptions and preferences regarding gamification vary based on several factors, including demographic differences and the alignment of game elements with users' goals. Research indicates that gamified services' success depends on the appeal of the game mechanics and how well these elements meet the educational and informational needs of different user groups (Marzal & Cardama, 2021).

### *User Attitudes Towards Gamification*

Users generally perceive gamification positively, especially in academic libraries, where it enhances engagement through playful and interactive methods. Studies have shown that gamification fosters active learning and interaction, making library services more enjoyable and accessible (Marzal & Cardama, 2021). For example, Jansukpum et al. (2024) found that users who engaged with gamified services, especially when combined with immersive technologies like Virtual Reality (VR), reported higher satisfaction and knowledge retention levels.

When effectively aligned with user goals, gamification enhances the learning experience by introducing elements like points, leaderboards, and rewards. These game mechanics increase user participation and motivate users to continue engaging with library services over time (Jansukpum et al., 2024). However, it is crucial to avoid overwhelming users with irrelevant or overly complex game mechanics, which could lead to disengagement (Marzal & Cardama, 2021).

Research also indicates that user excitement about gamified services may diminish unless new challenges and features are introduced to maintain engagement. For example, Chernbumroong et al. (2024) found that while initial excitement about gamified library systems like the Angkeaw Library Automation was high, user engagement declined as the novelty wore off, underscoring the need for continuous updates and re-engagement strategies.

### *Demographic Differences*

Gamification appeals differently to various user demographics, with younger users, particularly undergraduate students, generally responding more positively to gamified services. This group tends to be more familiar with digital technologies and interactive learning environments, which makes them more receptive to game-like features (Chernbumroong et al., 2024). Younger users often value competitive elements such as points and leaderboards, making their learning experiences more engaging and enjoyable (Reed & Miller, 2020).

In contrast, older users, such as postgraduate students and faculty members, may be less enthusiastic about gamified systems, often viewing them as less relevant to their academic or professional needs. Faculty members, for instance, may prioritize library services' efficiency and practical utility over the entertainment aspects of gamification (Reed & Miller, 2020). Postgraduate students engage more with gamified services if they directly support skill development or professional growth, as demonstrated by Marzal & Cardama's (2021) research on gamification in visual literacy courses.

Additionally, the appeal of gamification can vary based on academic discipline. For example, students in humanities may appreciate creative and narrative-driven gamified elements, while STEM students may prefer gamified tools that emphasize problem-solving and data analysis (Marzal & Cardama, 2021). These differences highlight the importance of tailoring gamification strategies to meet the diverse needs of different user groups.

Research by Capdarest-Arest et al. (2019) further illustrates that professional demographics, such as health sciences information professionals, can also benefit from gamification. In a study involving a gamification workshop, participants reported increased confidence in using gamified tools, suggesting that even professionals with limited prior exposure to gamification can be positively influenced by hands-on experience. The workshop results showed that 86% of participants planned to incorporate gamification into their instructional practices, demonstrating the broad potential of gamification across various professional settings (Capdarest-Arest et al., 2019).

While gamification can potentially engage a wide range of users, its success depends on understanding and addressing different demographic groups' preferences and motivations. Libraries must carefully design gamified services to ensure they remain relevant, enjoyable, and aligned with users' academic and informational needs across diverse populations.

*Research Methodology*
*3.1 Research Design*
This study employs a quantitative research approach, utilizing a survey-based methodology to examine the role of gamification in enhancing academic library services. The survey is designed with structured, closed-ended questions to gather comprehensive data from library leaders across various educational institutions in India.

I. The questionnaire consists of five main sections, each addressing key areas:
II. Demographic Information – capturing respondents' roles, experience, and institutional affiliations.
III. Awareness of Gamification – exploring knowledge of gamification concepts and tools.
IV. Perceptions of Gamification – examining beliefs about the effectiveness of gamification in libraries.
V. Challenges to Implementation – identifying barriers faced by libraries.
VI. Resource Requirements – determining what is needed to support gamification efforts.

The structured survey ensures consistency in responses, making it suitable for broad statistical analysis while providing detailed insights into the current state of gamification in academic libraries in India.

*3.2 Sampling*
The sampling method used is stratified random sampling. This approach ensures a representative sample of library leaders, including librarians, deputy librarians, and assistant librarians from college libraries, university libraries, and other educational institutions across India. The stratification accounts for variations in the leadership hierarchy, recognizing that assistant librarians may play critical roles in library management in some institutions.

Participants' email addresses were obtained from institutional directories, and the survey was distributed through Google Forms. To increase the response rate, the survey link was also shared in official WhatsApp groups of the Indian Library Association. A total of 700 invitations were sent, and after data cleaning, 273 usable responses were retained for analysis. This sample size ensures a wide-ranging view of library leadership perspectives on gamification.

### *3.3 Data Collection*
The survey was primarily administered via Google Forms, enabling efficient data management and collection. A pilot study was conducted to test the survey's clarity and structure, involving a small group of library professionals. Based on their feedback, modifications were made to improve the relevance and precision of the questions.

Of the 700 surveys distributed, 280 responses were initially received. Following data cleaning—which involved removing incomplete or duplicate entries and ensuring only relevant participants were included—273 valid responses were retained. This final sample reflects a broad range of institutions and geographic diversity, representing library leaders from various regions of India, providing a solid foundation for analysis.

### *3.4 Data Analysis*
Data analysis was performed using SPSS (Statistical Package for the Social Sciences). The analysis included:
- Descriptive Statistics: Calculations of means, medians, modes, frequencies, and percentages to summarize the data and identify overall trends.
- Inferential Statistics: A t-test for independent means was conducted to examine potential differences in awareness and perception based on gender. A significance level of $p < 0.05$ was used for all tests, employing a two-tailed test to identify statistically significant differences.

These analytical methods helped uncover trends, correlations, and significant findings related to the impact of gamification on library services. Figures and charts were created using Excel and Python to visualize the findings, ensuring clarity and accessibility when presenting the results.

### *3.5 Ethical Considerations*
All ethical considerations were carefully observed throughout the study. Informed consent was obtained from participants before they completed the survey. Respondents were informed about the study's purpose, assured of their right to withdraw at any stage, and guaranteed anonymity. All personal data was handled securely, and responses were used solely for this research.

### *Result*
Figure 1: Demographic Information of Respondents

Most respondents are librarians or chief library officers (81.3%), suggesting a strong representation of senior library leadership. A significant proportion are male (66.3%), and most have 11-15 years of experience (28.2%). Most respondents are from colleges (51.3%) and located in Western India (35.2%). This demographic distribution provides a comprehensive view of library leadership in India, reflecting diverse experience levels and institutional types.

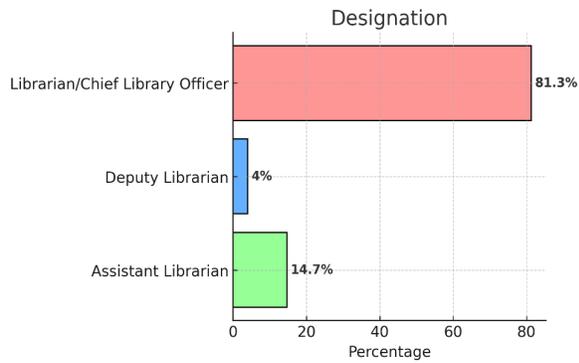
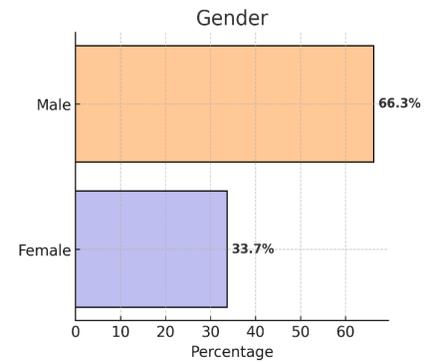
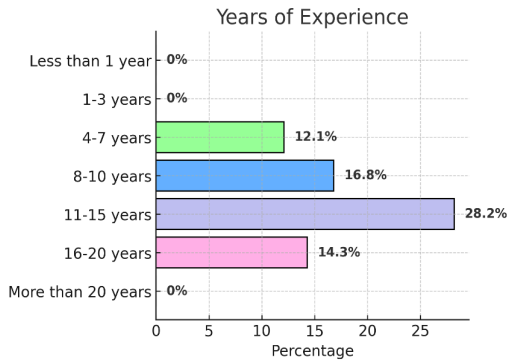
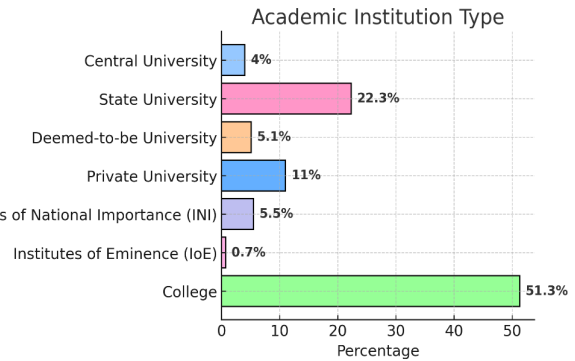
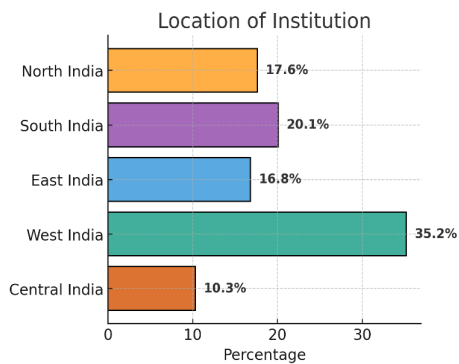

Table 1: Awareness of Gamification Among Library Leaders

The results show a moderate level of awareness among library leaders regarding gamification, with the highest awareness of its benefits for user engagement (mean = 3.36) and the concept of gamification itself (mean = 3.32). However, awareness of specific elements, such as different types of gamification tools and successful examples, is somewhat lower. The significant t-values and low p-values across the variables indicate that these findings are statistically significant, highlighting a need for increased awareness and training in gamification strategies.

| Awareness Areas | Mean | Median | Mode | SD | T-Value |
|---|---|---|---|---|---|
| Concept of gamification in academic libraries | 3.32 | 4 | 4 | 1.13 | -24.08 |
| Different types of gamification elements (points, badges, leaderboards) | 3.17 | 3 | 4 | 1.11 | -22.95 |

| | Mean | Median | Mode | SD | T-Value |
|---|---|---|---|---|---|
| Benefits of gamification in enhancing user engagement | 3.36 | 4 | 4 | 1.13 | -24.48 |
| Challenges and limitations of implementing gamification | 3.28 | 3 | 4 | 1.13 | -23.54 |
| Examples of successful gamification implementations in academic libraries | 3.09 | 3 | 3 | 1.13 | -21.16 |
| Technological tools and platforms supporting gamification | 3.16 | 3 | 4 | 1.10 | -23.14 |
| Process of designing a gamification strategy for library services | 3.03 | 3 | 4 | 1.16 | -19.67 |
| Steps involved in integrating gamification elements into existing library programs | 3.08 | 3 | 4 | 1.16 | -20.22 |
| Ethical considerations and data privacy issues related to gamification | 3.02 | 3 | 3 | 1.11 | -20.81 |
| Potential for gamification to improve learning outcomes and information literacy | 3.19 | 3 | 4 | 1.15 | -21.67 |

The Scale Used: Very much aware (5) Aware (4) Somewhat aware (3) Slightly aware (2) Not aware (1)
The result is significant at p < .05 and P-Value < .00001.

Table 2: Perceptions of Gamification in Academic Libraries

Perceptions of gamification in academic libraries are generally neutral, with means around 2.83 to 3.01, suggesting that respondents are neither strongly in favor nor against gamification. Key areas of neutral to slightly positive perception include its potential to improve learning experiences and increase engagement, though views on its effectiveness for promoting information literacy and retention are less positive. The significant t-values and p-values indicate that these perceptions are statistically significant, reflecting a cautious but not overly enthusiastic stance towards gamification.

| Perceptions of Gamification in Academic Libraries (To what extent do you agree with the statement?) | Mean | Median | Mode | SD | T-Value |
|---|---|---|---|---|---|
| Gamification can improve the overall learning experience for library users | 2.87 | 3 | 4 | 1.14 | -18.34 |
| Gamification can increase student engagement with library resources | 2.96 | 4 | 4 | 1.12 | -19.79 |
| Gamification encourages more frequent visits to the library | 3.01 | 4 | 4 | 1.12 | -20.57 |
| Gamification can effectively promote information literacy skills among library users | 2.84 | 3 | 4 | 1.15 | -17.59 |
| Gamification helps create a more interactive and enjoyable library environment | 2.83 | 3 | 4 | 1.15 | -17.51 |
| Gamification can lead to better retention of library patrons | 2.73 | 3 | 4 | 1.14 | -16.66 |
| The use of gamification in libraries can foster a sense of community among users | 2.91 | 3 | 4 | 1.11 | -19.76 |
| Gamification can be an effective tool for marketing library services and resources | 2.85 | 3 | 4 | 1.13 | -18.14 |
| Gamification is a valuable method to encourage collaborative learning in the library | 2.97 | 3 | 4 | 1.10 | -20.61 |
| Gamification strategies in libraries can positively impact academic success | 2.76 | 3 | 4 | 1.13 | -17.1 |

The Scale Used: Strongly agree (5) Agree (4) Neutral (3) Disagree (2) Strongly disagree (1)
The result is significant at $p < .05$ and P-Value $< .00001$.

Figure 2: Library Services That Could Benefit from Gamification
Library services such as information literacy sessions (74.7%), library tours and orientations (71.1%), and online resources and database usage (67.4%) are identified as the most likely to benefit from gamification. The high percentages for these services suggest that library leaders see significant potential for gamification to enhance these areas. Lower percentages for services like catalog searching and navigation (42.9%) indicate that not all services are perceived as equally amenable to gamification.

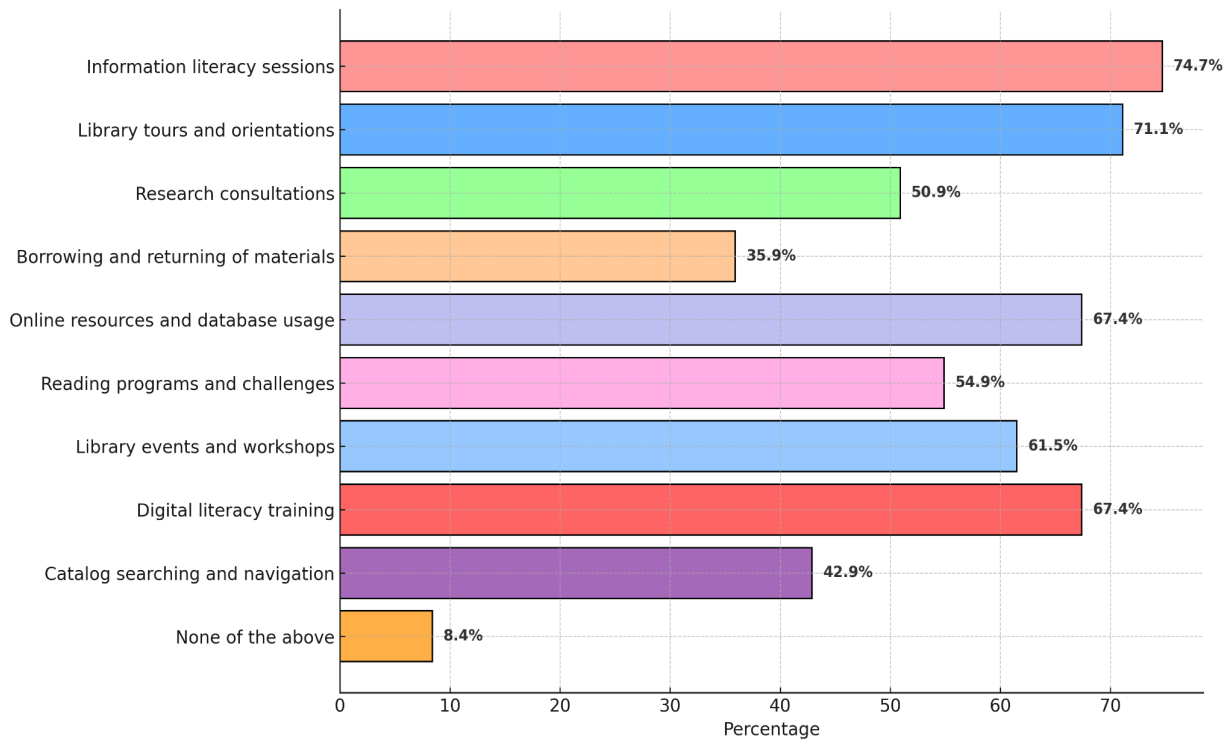

Figure 3: Demographic Appeal of Gamification
Gamification is perceived to appeal most to undergraduate students (49.1%) and postgraduate students (37%), with less emphasis on PhD students and faculty. A notable proportion of respondents (38.5%) believe gamification appeals equally to all demographics, while a small percentage (7.3%) feel it does not appeal to any specific group. This distribution suggests that while gamification may be particularly attractive to undergraduate and postgraduate students, there is also a recognition of its broader appeal.

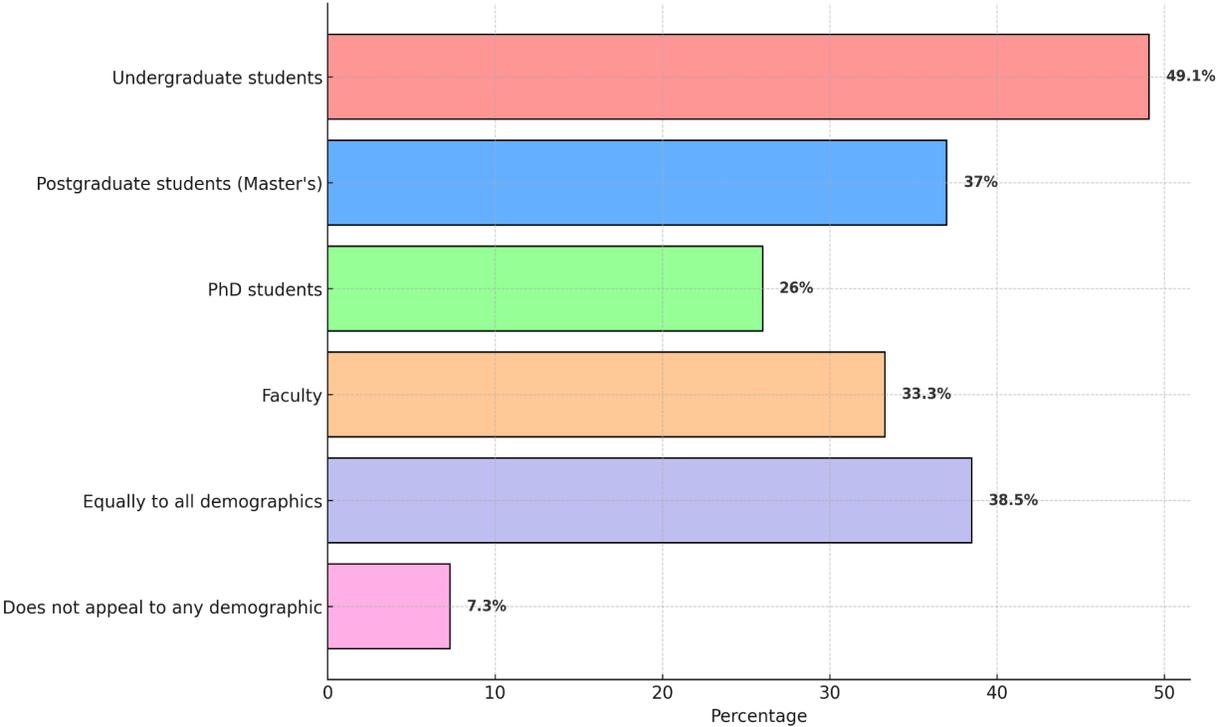

Figure 4: Challenges in Implementing Gamification

A substantial portion of libraries (44.3%) reports encountering challenges in implementing gamification strategies. This indicates that while the concept is recognized, practical barriers are significant. The remaining respondents either have not faced challenges (16.5%) or find the issue not applicable (37%), suggesting varying levels of experience and readiness for gamification.

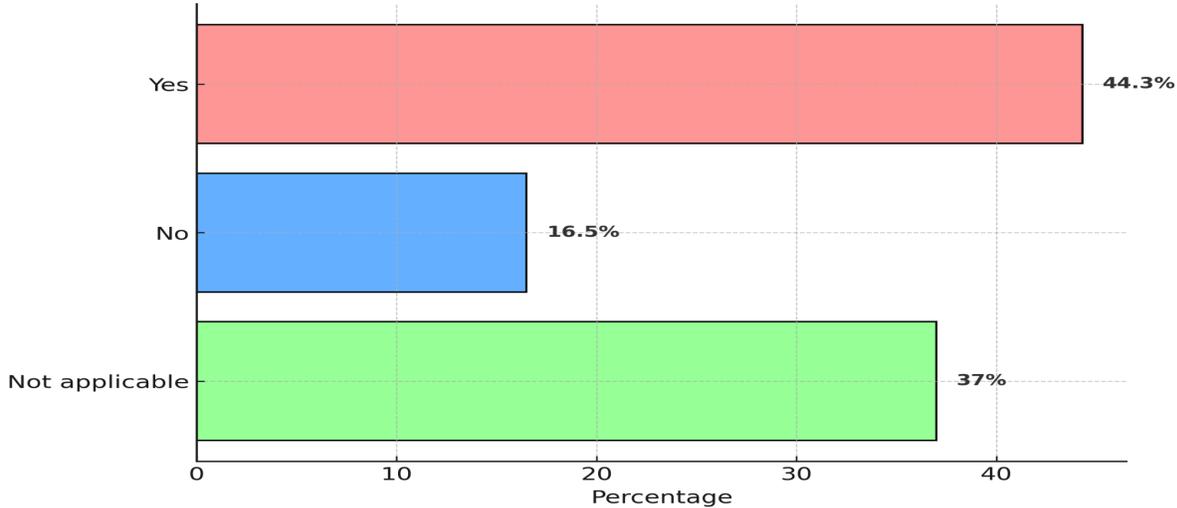

Figure 5: Primary Challenges in Implementing Gamification

The primary challenges identified include limited staff expertise (50.2%), insufficient technological infrastructure (42.5%), and lack of funding (40.7%). These challenges highlight the need for better resources, training, and support to overcome obstacles in gamification implementation. Resistance to change and difficulties in integration also contribute to the complexity of adopting gamification in library settings.

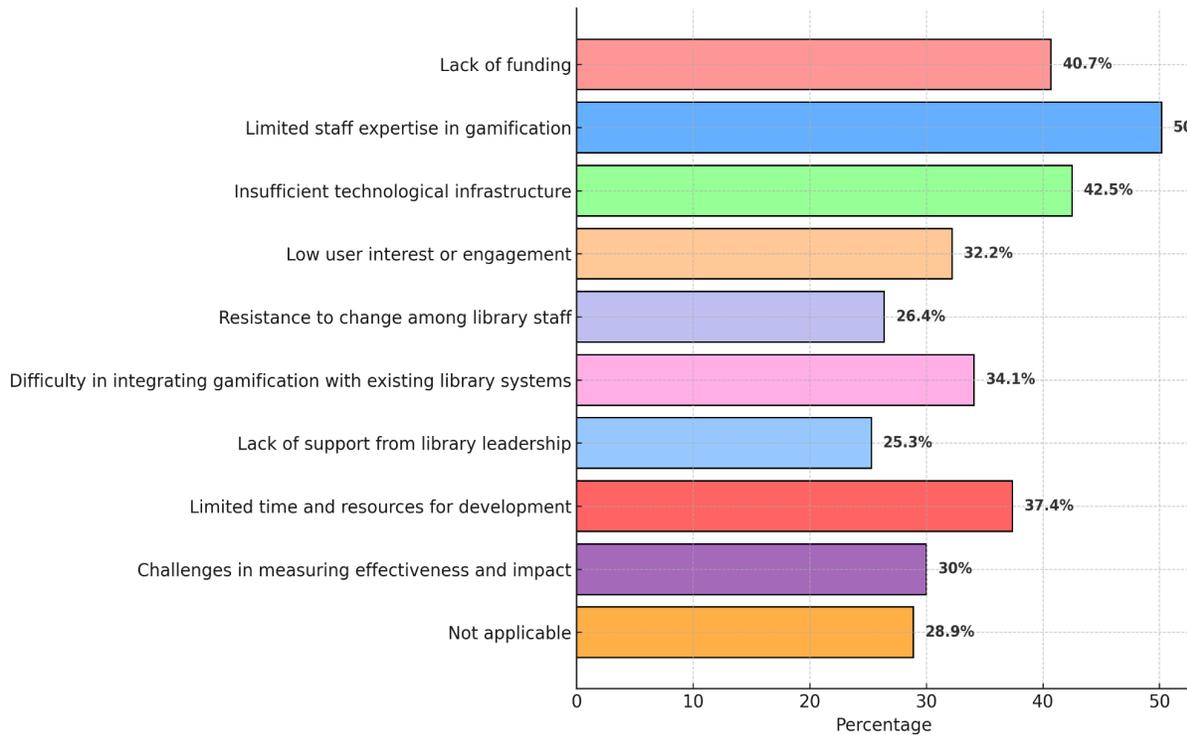

Figure 6: Additional Resources or Support for Gamification

Respondents identified several key resources needed for successful gamification, including access to innovative software (56.8%), increased staffing (53.8%), and training (54.9%). The high need for these resources suggests that libraries are aware of the potential of gamification but require additional support and infrastructure to fully realize its benefits. Clear policy guidelines and technical support are also emphasized as critical for effective implementation.

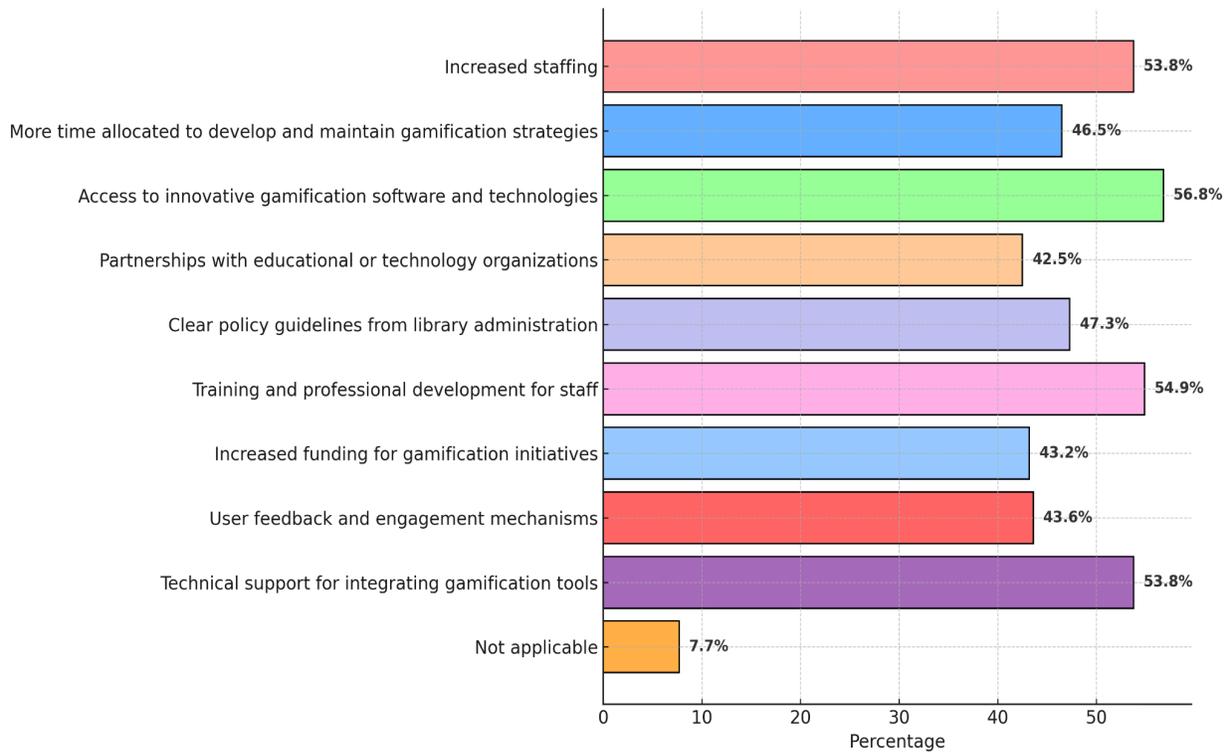

*Discussion*

This study examines the role of gamification in enhancing academic library services in India. The findings indicate that while gamification is recognized for its potential to improve user engagement, information literacy, and research participation, challenges related to implementation persist. These challenges include limited staff expertise, insufficient technological infrastructure, and funding constraints. Addressing these issues through policy support, training, and investment in digital tools can facilitate successful gamification integration.

*Awareness and Perceptions of Gamification*

The study highlights moderate awareness of gamification among library leaders, with respondents demonstrating a better understanding of its potential benefits in user engagement (Mean = 3.36) and overall learning experience (Mean = 3.32). However, there is less awareness regarding technological tools and strategies required for implementation. This aligns with previous studies suggesting that while the concept of gamification is familiar, its practical application remains underutilized due to limited knowledge and training (Sánchez-Mena & Martí-Parreño, 2017). Libraries must bridge this gap by offering professional development programs focused on gamification strategies and technology integration.

*Potential Applications in Library Services*

Gamification is perceived as most beneficial for information literacy sessions (74.7%), library tours and orientations (71.1%), and online resources and database usage (67.4%). This aligns with research suggesting that gamification can make information literacy programs more engaging and interactive (Jansukpum et al., 2024; Saikia et al., 2023). Interactive tools such as quizzes, digital escape rooms, and leaderboards can enhance participation and knowledge retention. However, services such as catalog searching and navigation (42.9%) received lower responses, indicating that not all library operations may benefit equally from gamification.

### *Demographic Appeal and User Engagement*
Findings suggest that gamification appeals most to undergraduate students (49.1%) and postgraduate students (37%), while PhD students and faculty show comparatively lower engagement levels. Younger users, being more accustomed to digital tools and interactive learning, respond positively to game-based approaches (Reed & Miller, 2020). Tailoring gamification strategies to different user demographics can ensure broader participation and sustained engagement.

### *Key Challenges in Gamification Implementation*
Several barriers hinder gamification adoption in academic libraries, with the most significant being limited staff expertise (50.2%), insufficient technological infrastructure (42.5%), and lack of funding (40.7%). These findings are consistent with studies highlighting the need for digital investment and professional training (Zeybek & Saygı, 2024). Resistance to change among library staff (26.4%) and challenges in measuring effectiveness (30%) were also reported, indicating that a structured training and impact assessment is necessary for long-term success.

### *Strategies for Overcoming Implementation Barriers*
To overcome these challenges, libraries should invest in professional training to equip staff with gamification knowledge and technical expertise. Cost-effective digital tools, such as open-source platforms like Kahoot! and ClassDojo, can be adopted to facilitate engagement without significant financial strain. Institutional support is crucial for securing funding and policy backing, ensuring gamification becomes a strategic initiative rather than a short-term experiment. Designing user-centric gamification experiences aligned with user needs will enhance adoption, while regular impact assessments using performance metrics and feedback mechanisms will help refine strategies for sustained success.

### *Implications for Future Research and Practice*
Future studies should focus on evaluating the long-term impact of gamification on library engagement and academic success. Comparative research across different library settings and user demographics can provide insights into optimizing gamification strategies. Moreover, understanding the correlation between awareness levels and perceived barriers can help in designing targeted interventions for improved gamification adoption.

### *Conclusion*
Gamification holds promise in transforming academic library services in India by enhancing user engagement and information literacy. However, addressing challenges such as staff expertise, funding, and technological constraints is crucial for its success. By adopting strategic policies, investing in training, and leveraging cost-effective digital tools, academic libraries can create more engaging and interactive learning environments. Future research should explore structured gamification models tailored to library settings and assess their effectiveness in fostering long-term academic engagement and research participation.

**Acknowledgement**
Naved Ahmad would like to thank AlMaarefa University, Riyadh, Saudi Arabia, for supporting this research.

**Author agreement/declaration**
This is a statement to certify that all authors have seen and approved the final version of the manuscript being submitted. This manuscript has not received prior publication and is not under consideration for publication elsewhere.

**Funding**
This research did not receive any specific grant from funding agencies in the public, commercial, or not-for-profit sectors.

**Declaration of competing interest**
None.

*References*

Adeyemi, I. O., Esan, A. O., & Aleem, A. (2021). Application of gamification to library services: Awareness, perception, and readiness of academic librarians in Nigeria. *The Electronic Library*, *39*(5), 767–781. https://doi.org/10.1108/EL-05-2021-0096

Aibar-Almazán, A., Castellote-Caballero, Y., Carcelén-Fraile, M. del C., Rivas-Campo, Y., & González-Martín, A. M. (2024). Gamification in the classroom: Kahoot! As a tool for university teaching innovation. *Frontiers in Psychology*, *15*. https://doi.org/10.3389/fpsyg.2024.1370084

Alaban, A., & Singh, S. B. (2024). Gamification Techniques to Enhance Student Engagement in Library Services. *Library Progress International*, *44*(1s), Article 1s.

Alsawaier, R. S. (2018). The effect of gamification on motivation and engagement. *The International Journal of Information and Learning Technology*, *35*(1), 56–79. https://doi.org/10.1108/IJILT-02-2017-0009

Appleton, L. (2020). Academic Libraries and Student Engagement: A Literature Review. *New Review of Academic Librarianship*, *26*(2–4), 189–213. https://doi.org/10.1080/13614533.2020.1784762

Brigham, T. J. (2015). An Introduction to Gamification: Adding Game Elements for Engagement. *Medical Reference Services Quarterly*, *34*(4), 471–480. https://doi.org/10.1080/02763869.2015.1082385

Brown, R. T., & Kasper, T. (2013). The Fusion of Literacy and Games: A Case Study in Assessing the Goals of a Library Video Game Program. *Library Trends*, *61*(4), 755–778.

Cai, Y., Li, X., & Shi, W. (2024). Does gamification affect knowledge-sharing behavior? The mediating role of intrinsic satisfaction needs. *Online Information Review*, *48*(2), 354–373. https://doi.org/10.1108/OIR-05-2021-0288

Capdarest-Arest, N., Opuda, E., & Stark, R. K. (2019). "Game on!" Teaching gamification principles for library instruction to health sciences information professionals using interactive, low-tech activities and design-thinking modalities. *Journal of the Medical Library Association*, *107*(4), Article 4. https://doi.org/10.5195/jmla.2019.636

Chernbumroong, S., Niemsup, S., Worragin, P., Homla, P., & Puritat, K. (2024). Enhancing User Engagement and Interaction in Library Automation through Gamification. *2024 Joint International Conference on Digital Arts, Media and Technology with ECTI Northern Section Conference on Electrical, Electronics, Computer and Telecommunications Engineering (ECTI DAMT & NCON)*, 270–274.

Colasanti, N., Fiori, V., & Frondizi, R. (2020). Promoting knowledge circulation in public libraries: The role of gamification. *Library Management*, *41*(8/9), 669–676. https://doi.org/10.1108/LM-04-2020-0064

Cziksentmihalyi, M. (1990). *Flow – The Psychology of optimal experience*. Harper & Row.

Dalili Saleh, M., Salami, M., Soheili, F., & Ziaei, S. (2022). Augmented reality technology in the libraries of universities of medical sciences: Identifying the application, advantages and challenges and presenting a model. *Library Hi Tech*, *40*(6), 1782–1795. https://doi.org/10.1108/LHT-01-2021-0033

Deterding, S., Dixon, D., Khaled, R., & Nacke, L. (2011). From game design elements to gamefulness: Defining "gamification." *Proceedings of the 15th International Academic MindTrek Conference: Envisioning Future Media Environments*, 9–15. https://doi.org/10.1145/2181037.2181040

Eden, B. L. (2015). *Enhancing Teaching and Learning in the 21st-Century Academic Library: Successful Innovations That Make a Difference*. Rowman & Littlefield.

Encheva, M., Tammaro, A. M., & Kumanova, A. (2020). Games to Improve Students Information Literacy Skills. *The International Information & Library Review*, *52*(2), 130–138.

Felker, K., & Phetteplace, E. (2014). Gamification in Libraries: The State of the Art. *Reference & User Services Quarterly*, *54*(2), 19–23.

Folmar, D. (2015). *Game It Up!: Using Gamification to Incentivize Your Library*. Rowman & Littlefield.

Furferi, R., Di Angelo, L., Bertini, M., Mazzanti, P., De Vecchis, K., & Biffi, M. (2024). Enhancing traditional museum fruition: Current state and emerging tendencies. *Heritage Science*, *12*(1), 20. https://doi.org/10.1186/s40494-024-01139-y

Haasio, A., Madge, O.-L., & Harviainen, J. T. (2021). Games, Gamification and Libraries. In O.-L. Madge (Ed.), *New Trends and Challenges in Information Science and Information Seeking Behaviour* (pp. 127–137). Springer International Publishing. https://doi.org/10.1007/978-3-030-68466-2_10

Hamari, J., Koivisto, J., & Sarsa, H. (2014). Does Gamification Work? – A Literature Review of Empirical Studies on Gamification. *2014 47th Hawaii International Conference on System Sciences*, 3025–3034. https://doi.org/10.1109/HICSS.2014.377

Heffernan, K. L., & Chartier, S. (2020). Augmented Reality Gamifies the Library: A Ride Through the Technological Frontier. In *Emerging Trends and Impacts of the Internet of Things in Libraries* (pp. 194–210). IGI Global. https://doi.org/10.4018/978-1-7998-4742-7.ch011

Hill, V., & Knutzen, K. B. (2017). Virtual world global collaboration: An educational quest. *Information and Learning Science*, *118*(9/10), 547–565. https://doi.org/10.1108/ILS-02-2017-0010

Jansukpum, K., Chernbumroong, S., Intawong, K., Sureephong, P., & Puritat, K. (2024). Gamified Virtual Reality for Library Services: The Effect of Gamification on

Enhancing Knowledge Retention and User Engagement. *New Review of Academic Librarianship*, *0*(0), 1–21. https://doi.org/10.1080/13614533.2024.2381509

Jug, T., Balog, K. P., & Faletar, S. (2023). Exploring the role of games and gamification in academic libraries from the perspective of LIS educators. *Education for Information*, *39*(3), 341–358. https://doi.org/10.3233/EFI-230038

Kabilan, M. K., Annamalai, N., & Chuah, K.-M. (2023). Practices, purposes and challenges in integrating gamification using technology: A mixed-methods study on university academics. *Education and Information Technologies*, *28*(11), 14249–14281. https://doi.org/10.1007/s10639-023-11723-7

Khademi Zare, S., Varnaseri, A., & Bayati, N. (2024). Identifying the Applications of Gamification for Audience Attraction in Public Libraries. *Libri*, *74*(2), 171–184. https://doi.org/10.1515/libri-2023-0123

Kharat, S. A., Nagarkar, S., & Panage, B. (2024). Information consolidation and repackaging for augmented reality library service: A special reference to the Layar app. *Information Discovery and Delivery*, *52*(1), 39–52. https://doi.org/10.1108/IDD-03-2022-0022

Kim, B. (2015). Chapter 4. Gamification in Education and Libraries. *Library Technology Reports*, *51*(2), Article 2.

Kitamura, T., & Sumi, Y. (2023). Using Gamification to Activate University Library Use. In X. Fang (Ed.), *HCI in Games* (pp. 343–359). Springer Nature Switzerland. https://doi.org/10.1007/978-3-031-35930-9_23

Latorre-Cosculluela, C., Sierra-Sánchez, V., & Vázquez-Toledo, S. (2025). Gamification, collaborative learning and transversal competences: Analysis of academic performance and students' perceptions. *Smart Learning Environments*, *12*(1), 2. https://doi.org/10.1186/s40561-024-00361-2

Leenaraj, B., Arayaphan, W., Intawong, K., & Puritat, K. (2023). A gamified mobile application for first-year student orientation to promote library services. *Journal of Librarianship and Information Science*, *55*(1), 137–150. https://doi.org/10.1177/09610006211067273

Lim, S.-K. (2023). On the Application of Gamification Elements in Libraries. *Journal of Information Science Theory and Practice*, *11*(2), 11. https://doi.org/10.1633/JISTaP.2023.11.2.1

Lischer-Katz, Z., & Cook, M. (2022). Virtual Reality and the Academic Library of the Future. *Transactions of the American Philosophical Society*, *110*(3), 185–210.

Malone, D. (2018, January 3). Escaping Library Orientation: The Introduction of Escape Rooms Into First-Year Experience Courses for Library Orientation and Familiarization [Journal]. *The Journal of Creative Library Practice*. https://creativelibrarypractice.org/2018/01/03/escaping-library-orientation-the-introduction-of-escape-rooms-into-first-year-experience-courses-for-library-orientation-and-familiarization/

Marzal, M.-Á., & Cardama, S. M. (2022). Gamification as a Strategy for Visual Literacy Skills-Based Education: A Proposal for Educational Libraries. *Journal of Library & Information Services in Distance Learning*, *15*(4), 236–252. https://doi.org/10.1080/1533290X.2021.2005215

Pinto, M., Garcia-Marco, J., Caballero, D., Manso, R., Uribe, A., & Gomez, C. (2024). Assessing information, media and data literacy in academic libraries: Approaches and challenges in the research literature on the topic. *The Journal of Academic Librarianship*, *50*(5), 102920. https://doi.org/10.1016/j.acalib.2024.102920

Reed, K., & Miller, A. (2020). Applying Gamification to the Library Orientation: A Study of Interactive User Experience and Engagement Preferences. *Information Technology and Libraries*, *39*(3), Article 3. https://doi.org/10.6017/ital.v39i3.12209

Richter, G., Raban, D. R., & Rafaeli, S. (2015). Studying Gamification: The Effect of Rewards and Incentives on Motivation. In T. Reiners & L. C. Wood (Eds.), *Gamification in Education and Business* (pp. 21–46). Springer International Publishing. https://doi.org/10.1007/978-3-319-10208-5_2

Riedmann, A., Schaper, P., & Lugrin, B. (2024). Integration of a social robot and gamification in adult learning and effects on motivation, engagement and performance. *AI & SOCIETY*, *39*(1), 369–388. https://doi.org/10.1007/s00146-022-01514-y

Rojas-Alfaro, R. (2024). Navigating the stacks virtually: Integrating virtual reality into writing resource instruction. *Computers and Composition*, *72*, 102851. https://doi.org/10.1016/j.compcom.2024.102851

Saikia, S., Gul, S., & Verma, M. K. (2023). Are libraries ready to serve gamification tools for teaching and learning? A review based on computational mapping. *Global Knowledge, Memory and Communication*, *ahead-of-print*(ahead-of-print). https://doi.org/10.1108/GKMC-04-2023-0114

Sanchez, E., van Oostendorp, H., Fijnheer, J. D., & Lavoué, E. (2020). Gamification. In A. Tatnall (Ed.), *Encyclopedia of Education and Information Technologies* (pp. 816–827). Springer International Publishing. https://doi.org/10.1007/978-3-030-10576-1_38

Sánchez-Mena, A., & Martí-Parreño, J. (2017). Drivers and Barriers to Adopting Gamification: Teachers' Perspectives. *Electronic Journal of E-Learning*, *15*(5), Article 5.

Sewell, A. (2021). Creating a Choose-Your-Own-Adventure Library Orientation: The Process of Using a Text-Based, Interactive Storytelling Tool to Take Orientation Virtual. *Journal of New Librarianship*, *6*, 23.

Skinner, B. F. (1965). *Science And Human Behavior*. Free Press.

Suresh Babu, S., & Dhakshina Moorthy, A. (2024). Application of artificial intelligence in adaptation of gamification in education: A literature review. *Computer Applications in Engineering Education*, *32*(1), e22683. https://doi.org/10.1002/cae.22683

Sylvia, C. (2021). Summer Reading 2021 is here! | Attleboro Public Library. *Summer Reading 2021 Is Here!* https://attleborolibrary.org/2021/06/summer-reading-2021-is-here/

Tang, Y. (2021). Help first-year college students to learn their library through an augmented reality game. *The Journal of Academic Librarianship*, *47*(1), 102294. https://doi.org/10.1016/j.acalib.2020.102294

teachingexpertise.com. (2024, April 2). *20 Library Activities For Elementary Students: Setup, Execution, And Tips - Teaching Expertise*. https://www.teachingexpertise.com/k-5/library-activities-for-elementary-students/

van Roy, R., & Zaman, B. (2017). Why Gamification Fails in Education and How to Make It Successful: Introducing Nine Gamification Heuristics Based on Self-Determination Theory. In M. Ma & A. Oikonomou (Eds.), *Serious Games and Edutainment Applications: Volume II* (pp. 485–509). Springer International Publishing. https://doi.org/10.1007/978-3-319-51645-5_22

Walsh, A. (2015). Playful Information Literacy: Play and information Literacy in Higher Education. *Nordic Journal of Information Literacy in Higher Education*, *7*(1), 80–94. https://doi.org/10.15845/noril.v7i1.223

Withorn, T., Eslami, J., Lee, H., Clarke, M., Caffrey, C., Springfield, C., Ospina, D., Andora, A., Castañeda, A., Mitchell, A., Kimmitt, J. M., Vermeer, W., & Haas, A. (2021). Library instruction and information literacy 2020. *Reference Services Review*, *49*(3/4), 329–418. https://doi.org/10.1108/RSR-07-2021-0046

Withorn, T., Messer Kimmitt, J., Caffrey, C., Andora, A., Springfield, C., Ospina, D., Clarke, M., Martinez, G., Castañeda, A., Haas, A., & Vermeer, W. (2020). Library instruction and information literacy 2019. *Reference Services Review*, *48*(4), 601–682. https://doi.org/10.1108/RSR-08-2020-0057

Young, J. (2016). Can Library Research Be Fun? Using Games for Information Literacy Instruction in Higher Education. *Georgia Library Quarterly*, *53*(3), 7. https://doi.org/10.62915/2157-0396.1973

Zare, S. K., Varnaseri, A., & Bayati, N. (2024). Identifying the Applications of Gamification for Audience Attraction in Public Libraries. *Libri*, *74*(2), 171–184. https://doi.org/10.1515/libri-2023-0123

Zeybek, N., & Saygı, E. (2024). Gamification in Education: Why, Where, When, and How?—A Systematic Review. *Games and Culture*, *19*(2), 237–264. https://doi.org/10.1177/15554120231158625

Zhou, L., & Yang, Y. (2023). Investigating gamification services of university libraries in China. *Digital Library Perspectives*, *39*(3), 326–337. https://doi.org/10.1108/DLP-02-2023-0019